\documentclass[fleqn,twoside]{article}
\usepackage{espcrc2}
\def\e{{\rm e}}
\def\ln{{\rm ln}}
\usepackage{graphicx}
\usepackage[figuresright]{rotating}

\newcommand{\AmS}{{\protect\the\textfont2
  A\kern-.1667em\lower.5ex\hbox{M}\kern-.125emS}}

\title{Fat and thin
       Fisher zeroes\thanks{%
This work was partially supported by EC IHP network grant
{\bf HPRN-CT-1999-00161:} ``EUROGRID''.
}
}

\author{W. Janke\address{Institut f\"ur Theoretische Physik,
        Universit\"at Leipzig, Augustusplatz 10/11,
        04109 Leipzig, Germany}, 
        D.A. Johnston\address[HW]{Department of Mathematics,
        Heriot-Watt University,
        Edinburgh, EH14 4AS, Scotland}
        and 
        M. Stathakopoulos\addressmark[HW]}
       
\begin{document}

\begin{abstract}
We show that it is possible to determine
the locus of Fisher zeroes in the thermodynamic limit
for the Ising model on planar (``fat'') $\phi^4$ random graphs and
their dual quadrangulations by matching
up the real part of the high- and low-temperature branches of
the expression for the free energy. Similar methods work
for the mean-field model on generic, ``thin'' graphs. 
Series expansions are very easy to obtain for such random graph
Ising models.
\vspace{1pc}
\end{abstract}

\maketitle

\section{INTRODUCTION}

One of the more remarkable results to emerge from the study
of various statistical mechanical models coupled to
two-dimensional quantum gravity is a solution of the Ising
model in field \cite{KBK}. In discrete form the coupling to gravity
takes the form of the spin models living on an annealed ensemble
of triangulations or quadrangulations, or their dual planar graphs.
The partition function for the Ising model on a single graph
$G^n$ with $n$ vertices
\begin{equation}
Z_{{\rm single}}(G^n,\beta,h) =
\sum_{\{\sigma\}} \e^{{\beta}\sum_{\langle i,j \rangle} \sigma_i
\sigma_j + h \sum_i \sigma_i}
\end{equation}
is promoted to a partition function which incorporates a sum
over some class of graphs ${\{G^n\}}$ by the coupling to gravity,
\begin{equation}
Z_n(\beta,h) = \sum_{\{G^n\}} Z_{{\rm single}}(G^n,\beta,h)\, .
\end{equation}
The solution to the Ising model in \cite{KBK} proceeded
by first forming the grand canonical partition function
\begin{equation}
{\cal Z} = \sum_{n=1}^{\infty} \left( - 4 g c \over ( 1 - c^2)^2 \right)^n
Z_n(\beta,h)
\label{grand}
\end{equation}
and then noting that this could be expressed as the free energy
\begin{eqnarray}
{\cal Z} \!\! &=& \!\!- \log \int {\cal D}\phi_1~{\cal D}\phi_2~ 
\exp \left( -{\rm Tr}\left[{1\over 2}(\phi_1^2+\phi_2^2) \right. \right.
\nonumber \\ 
& & -\left. \left. c\phi_1\phi_2   
+  \frac{g}{4}( \e^{h} \phi_1^4 + \e^{-h} \phi_2^4)\right]  \right) 
\label{matint}
\end{eqnarray}
of a matrix model, where we have written the potential that generates
$\phi^4$ graphs.
In the above $\phi_{1,2}$ are $N \times N$ Hermitian matrices,
$c = \exp ( - 2 \beta)$ and
the $N \to \infty$ limit is performed in order to pick out planar graphs.
The graphs of interest are generated as the Feynman diagrams of the
``action'' in equ.~(\ref{matint}), which is constructed so as to weight each
edge with the correct Boltzmann weights for nearest-neighbour interaction
Ising spins. Since the edges carry matrix indices the graphs in question
are ``fat'' or ribbon graphs.

The integral of equ.~(\ref{matint}) can be evaluated using the results
of \cite{mehta} to give (when $h=0$)
\begin{equation}
{\cal Z} \!=\! {1\over 2} \log \left( {z\over g} \right)
-{1\over g}\int_0^z\!\!{dt\over t}g(t)+{1\over 2g^2}
\int_0^z\!\!{dt\over t}g(t)^2,\,\,\, 
\label{fullpart}
\end{equation}
where $g$ is defined by
\begin{equation}
g(z)=3 c^2 z^3 +  z
\left[ \frac{1}{(1-3 z)^2}-c^2
\right].
\label{geq}
\end{equation}

\section{ZEROES}

The idea that 
the zeroes of the partition function
could determine the phase structure
of a spin model first appeared in
Lee and Yang's work \cite{LYM}.
They considered
how the non-analyticity characteristic of a phase transition
appeared from the partition function on finite
lattices or graphs, which was a polynomial
\begin{equation}
Z = \sum D_{mn} c^m y^n
\end{equation}
for a lattice with $m$ edges and $n$ vertices, again with
$c = \exp ( - 2 \beta), \, y = \exp ( -2 h )$.
They (and Fisher \cite{Fish}) showed that the behaviour of the zeroes of this polynomial
in the complex $y$ or $c$ plane,
in particular  the limiting locus as
$m,n \to \infty$,
determined the phase structure. 
For temperature driven transitions, 
in zero external field for simplicity, 
the thermodynamic limit 
of the free energy on some class of lattices or graphs $\{ G^n \}$ becomes
\begin{equation}
F(G^{\infty},\beta) \sim - \int_L d c \rho(c) \ln (c  - c(L) ),
\end{equation}
where $L$ is some set of curves, or possibly regions, in the
complex $c$ plane on which the zeroes have support and
$\rho(c)$ is the density of the zeroes there.

The general question of how to extract the locus of zeroes
analytically has
been considered by various authors, notably Shrock
and collaborators \cite{lottsashrock} for Ising and Potts models.
It was first observed in \cite{IPZ}
that such loci could be thought of as Stokes
lines separating different regions of
asymptotic behaviour of the partition
function in the complex temperature
or field planes. 
More recently, the case of
models with first-order transitions
has been investigated by
Biskup {\em et al.\/} \cite{BKT} 
who showed 
that
the partition function of a 
statistical mechanical
model defined in a periodic volume $V$ which depends
on some complex parameter such as $c$ or $y$ can be written in terms of
complex functions $F_l$ describing $k$ different phases, 
where the various $F_l$ are the metastable
free energies per unit volume of the phases, 
$\Re F_l = F$ characterises
the free energy when phase $l$ is stable.
The zeroes of the partition function are then determined
from the solutions of the
equations
\begin{eqnarray}
\Re F_{l} =  \Re F_{m} < \Re F_{k}, \; \; \forall k \ne l,m , \nonumber \\
\beta V ( \Im  F_{l}  -  \Im   F_{m} ) = \pi \; {\rm mod} \; 2 \pi .
\label{master}
\end{eqnarray}
The equations (\ref{master})
are thus in perfect agreement with the idea that the loci
of zeroes should be Stokes lines, since the
zeroes of $Z$ lie on the complex phase coexistence
curves $\Re F_{l} = \Re F_{m}$ in the complex  parameter
plane.

The specific Biskup {\em et al.\/} results apply to models with 
first-order transitions, 
but we are interested here in an Ising model with a third-order
transition, so it might initially seem that
these results were inapplicable.
We are saved by the fact that when considered
in the complex temperature plane
the transition is continuous only at the
physical point itself (and possibly some other
finite set of points). 

\section{FAT (AND THIN) ZEROES}

The determination of the locus of Fisher zeroes
for the Ising model on random graphs
in the thermodynamic limit using
the ideas of the previous section turns out to be rather straightforward,
as we now describe.
Since we wish to match $\Re F$ between the various
solution branches to obtain the loci of Fisher zeroes and
$ F \sim \log ( g (c) )$
for the Ising model
on planar graphs, the equation which determines
the loci of zeroes in the thermodynamic limit is
\begin{equation}
\log | g_L (c) | = \log | g_{H_{i}} ( c) |,
\label{ecurve}
\end{equation}
where the low-temperature solution
$g_L (c)$ and the various 
high-temperature solutions $g_{H_{i}} ( c)$ are given by solving
$g' ( z ) = 0$ in equ.~(\ref{geq}) for $z$.

The resulting curve is shown in the $c$ plane in Fig.~1
where it can be seen  that in addition to the 
physical phase transition at $c=1/4$, an unphysical
transition with the same KPZ \cite{KPZ} exponents appears
at $c=-1/4$.
The interior of the curve is the ferromagnetic $FM$ region
and the exterior the paramagnetic $PM$
and unphysical ``$O$'' phases, separated by 
cuts on the imaginary axis
which we have not plotted.

\begin{figure}[htb]
\hspace*{1.6cm}\includegraphics[scale=0.35]{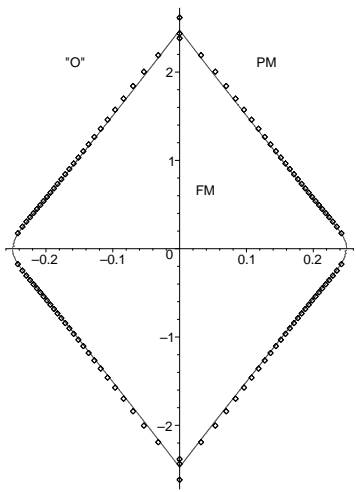}
\caption{The Fisher zeroes on fat $\phi^4$ graphs in the complex
$c$ plane.}
\label{fig1}
\end{figure}

The points plotted in Fig.~1 are generated from a series expansion of
${\cal Z}$ in equ.~(\ref{fullpart}), which is arrived at by reverting the
expression for $g(z)$ and substituting the resulting $z(g)$ into 
equ.~(\ref{fullpart}). 
Earlier work reported in \cite{jan} obtained similar results at lower
orders.
The form of the expression for ${\cal Z}$ means that the contributions
from each of the terms in equ.~(\ref{fullpart}) are proportional \cite{wjs}, 
so the full series for ${\cal Z}$ can be generated from $ 
{1\over 2} \log \left( z/g \right)$.

The loci of Fisher zeroes are highly non-universal, and we also show the 
zeroes on
``thin'', generic random $\phi^3$ graphs for comparison in Fig.~2.
These models can be thought of
as the $N \to 1$ limit of the matrix models, rather than the $N \to \infty$
planar limit.
Similar methods to those discussed above also serve 
in this case
where one has mean-field behaviour \cite{wjds}. 
For the Ising model on thin graphs $F \sim \log \tilde S$,
where $\tilde S$ is the saddle point action for either the
low- or high-temperature phase. The equivalent of equ.~(\ref{ecurve})
is then
\begin{equation}
| 2 ( 1 - c)^3 | = | ( 1 + c )^2 ( 1 - 2 c) |.
\end{equation}
Potts zeroes and chromatic zeroes are also accessible 
on the thin graphs. 

In summary, we have seen that an analytic determination
of Fisher zeroes for the Ising model on both fat and
thin random graphs is possible, and that series expansions
are easily obtainable.  The general form of the solution
also holds on (planar) random quadrangulations and $\phi^3$ graphs,
and in non-zero field, so all of these can also be investigated.
 
\begin{figure}[htb]
\hspace*{1.7cm}\includegraphics[scale=0.35]{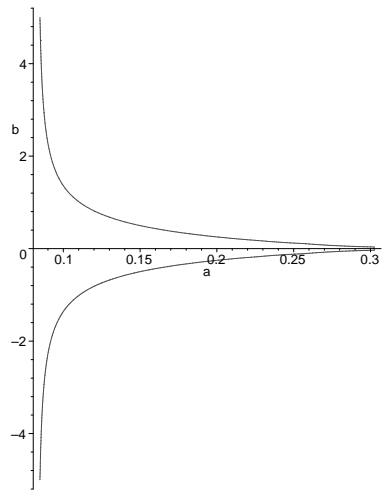}
\caption{The Fisher zeroes on thin $\phi^3$ graphs in the complex
$c$ plane.}
\label{fig2}
\end{figure}

\end{document}